\documentclass[12pt]{iopart}

\usepackage{graphicx}

\newcommand{\beq}{\begin{equation}}
\newcommand{\eeq}{\end{equation}}

\begin{document}

\title[Measurements of the fluctuation-induced magnetoconductivity in BaFe$_{2-x}$Ni$_x$As$_2$]{Measurements of the fluctuation-induced in-plane magnetoconductivity at high reduced temperatures and magnetic fields in the iron arsenide BaFe$_{2-x}$Ni$_x$As$_2$
}

\author{R I Rey$^1$, C Carballeira$^1$, J Mosqueira$^1$, S Salem-Sugui Jr.$^2$, A D Alvarenga$^3$, H -Q Luo$^4$, X -Y Lu$^4$, Y -C Chen$^4$ and F Vidal$^1$}

\address{$^1$ LBTS, Facultade de F\'isica, Universidade de Santiago de Compostela, E-15782 Santiago de Compostela, Spain}

\address{$^2$ Instituto de Fisica, Universidade Federal do Rio de Janeiro, 21941-972 Rio de Janeiro, RJ, Brazil}

\address{$^3$ Instituto Nacional de Metrologia Qualidade e Tecnologia, 25250-020 Duque de Caxias, RJ, Brazil}

\address{$^4$ Beijing National Laboratory for Condensed Matter Physics, Institute of Physics, Chinese Academy of Sciences, Beijing 100190, China}

\ead{j.mosqueira@usc.es}

\begin{abstract}
The superconducting fluctuations well inside the normal state of Fe-based superconductors were studied through measurements of the in-plane paraconductivity and magnetoconductivity in high quality BaFe$_{2-x}$Ni$_x$As$_2$ crystals with doping levels from the optimal ($x=0.10$) up to the highly overdoped ($x=0.20$). These measurements, performed in magnetic fields up to 9 T perpendicular to the \textit{ab} (Fe) layers, allowed a reliable check of the applicability to iron-based superconductors of Ginzburg-Landau approaches for 3D anisotropic compounds, even at high reduced temperatures and magnetic fields. Our results also allowed us to gain valuable insights into the dependence on the doping level of some central superconducting parameters (coherence lengths and anisotropy factor).
\end{abstract}

\maketitle

\section{Introduction}

As the pairing mechanism of Fe-based superconductors is not yet established, the phenomenological descriptions of their superconducting transition are at the forefront of the research in these materials.\cite{review} A powerful tool to probe these descriptions is the superconducting fluctuation effects that appear in the normal state.\cite{tesanovic,sugui,choi,mosqueira11,prando,flucC,pallecchi,fanfarillo,kim,putti,pandya} At present, some experimental aspects of these fluctuations have been studied through observables like the magnetization\cite{sugui,choi,mosqueira11,prando}, the specific heat\cite{flucC} or the electrical conductivity\cite{pallecchi,fanfarillo,kim,putti,pandya}. However, their behavior at high reduced temperatures and magnetic fields (in the short wavelength fluctuation regime) and their onset temperature, remains at present unexplored at a quantitative level. The importance of these aspects, which include the possible presence of phase fluctuations well above the superconducting critical temperature ($T_c$),\cite{sugui,choi,mosqueira11,prando} is enhanced by the comparison with the high-$T_c$ cuprate superconductors (HTSC), for which the  onset and the influence of doping on their superconducting fluctuations is at present one of the most central and debated issues of their phenomenology.\cite{parker,bilbro,kondo,rourke,ramallo12} 

The central aim of this paper is to present detailed experimental results on the superconducting fluctuations above $T_c$ in iron-based superconductors as a function of the doping level. To achieve this, we have measured  in an extended temperature region above the superconducting transition the in-plane fluctuation electric conductivity, $\Delta\sigma_{ab}(T,H)$, in optimally-doped and overdoped BaFe$_{2-x}$Ni$_x$As$_2$  crystals ($0.1\leq x\leq 0.2$), under magnetic fields up to 9 T perpendicular to the \textit{ab} (Fe) layers. We have also used these experimental data to probe the phenomenological descriptions of the superconducting fluctuations based on the Gaussian Ginzburg-Landau (GGL) approach, adapted to the 3D anisotropic nature of these compounds, and also to take into account the short wavelength fluctuation regime.\cite{PRBcarballeira,EPLcutoff} Our analysis allows us to gain valuable insight into the doping dependence of their superconducting parameters. In particular, the anisotropy factor is shown to increase significantly in strongly overdoped samples.

\section{Experimental details and results}

\subsection{Crystals fabrication and characterization}

The BaFe$_{2-x}$Ni$_x$As$_2$ samples used in this work are plate-like single crystals (typically $5\times2\times0.1$ mm$^3$) with the $c$ crystallographic axis perpendicular to their largest face. They were cleaved from larger single crystals grown by the self-flux method. Their nominal Ni doping levels are $x=0.10$, 0.15, 0.18 and 0.20, although the real doping level was found to be a factor $\sim0.8$ smaller (see Ref.~\cite{growth}, where all the details of the growth procedure and characterization may be found). We checked the excellent stoichiometric and structural quality of the crystals studied here by x-ray diffraction. In particular, the $(00l)$ linewidths were found to be slightly larger [$\Delta(2\theta)\sim0.10^\circ$, FWHM] than the corresponding instrumental linewidths, $\Delta(2\theta)\sim0.07^\circ$. This was attributed to a dispersion in the $c$-axis lattice parameter $L_c$,\footnote{A similar procedure was previously used to investigate compositional inhomogeneities in non-stoichiometric high-$T_c$ cuprates, see e.g., Ref.~\cite{intrinsic}} and was used to roughly estimate $\Delta x\sim10^{-2}$ through the $L_c(x)$ dependence presented in Ref.~\cite{growth}. In turn, the rocking curves for the (008) lines indicated that the dispersion in the $c$-axis orientation was lower than $\sim0.05^\circ$.

\subsection{Resistivity measurements: Determination of the superconducting transition temperatures and transition widths}

The in-plane resistivity (along the \textit{ab} layers) was measured in the presence of magnetic fields up to 9 T perpendicular to the $ab$ layers with a Quantum Design physical property measurement system (PPMS) by using four contacts with an in-line configuration and an excitation current of $\sim 1$~mA at 23 Hz. The uncertainties in the geometry and dimensions of the crystals, and the finite size of the electrical contacts (stripes typically 0.5 mm wide) lead to an uncertainty in the $\rho_{ab}$ amplitude of $\sim25$\%. An example of the $\rho_{ab}(T)$ dependence around the transition (corresponding to the crystal with $x$=0.1) is presented in Fig.~1(a). The zero-field transition temperature, $T_c$, was estimated from the maximum of the corresponding $d\rho_{ab}/dT$ curve, and the transition width as $\Delta T_c\approx 2(T_c-T_c^-)$, where $T_c^-$ is the temperature at which $\rho_{ab}=0$ is attained. In the optimally doped crystal ($x=0.1$) $T_c=20.0$~K and $\Delta T_c\approx0.3$~K, which is among the best in the literature for crystals of the same composition.\cite{tao,sun09,ni10,shahbazi11} This allows investigation of fluctuation effects down to reduced temperatures, $\varepsilon\equiv\ln(T/T_c)$, of the order of $\Delta T_c/T_c\approx 10^{-2}$. Overdoped crystals present a slightly wider resistive transition ($\Delta T_c\approx0.6$~K) probably due to the $T_c(x)$ dependence and an $x$-distribution (taking into account that in the overdoped regime $|dT_c/dx|\sim200$~K it may be estimated $\Delta x\sim0.003$). Nevertheless, even in these samples, fluctuation effects may be studied in a wide range of temperatures above the zero-field $T_c$ by just applying a magnetic field of the order of $\sim$1 T due to the $T_c(H)$ shift to lower temperatures. A detailed analysis of the effect on the measured fluctuation conductivity of the uncertainty in $T_c$ is presented in Appendix A.

\subsection{Determination of the fluctuation contribution to the electric conductivity}

\begin{figure}[t]
\begin{center}
\includegraphics[scale=.8]{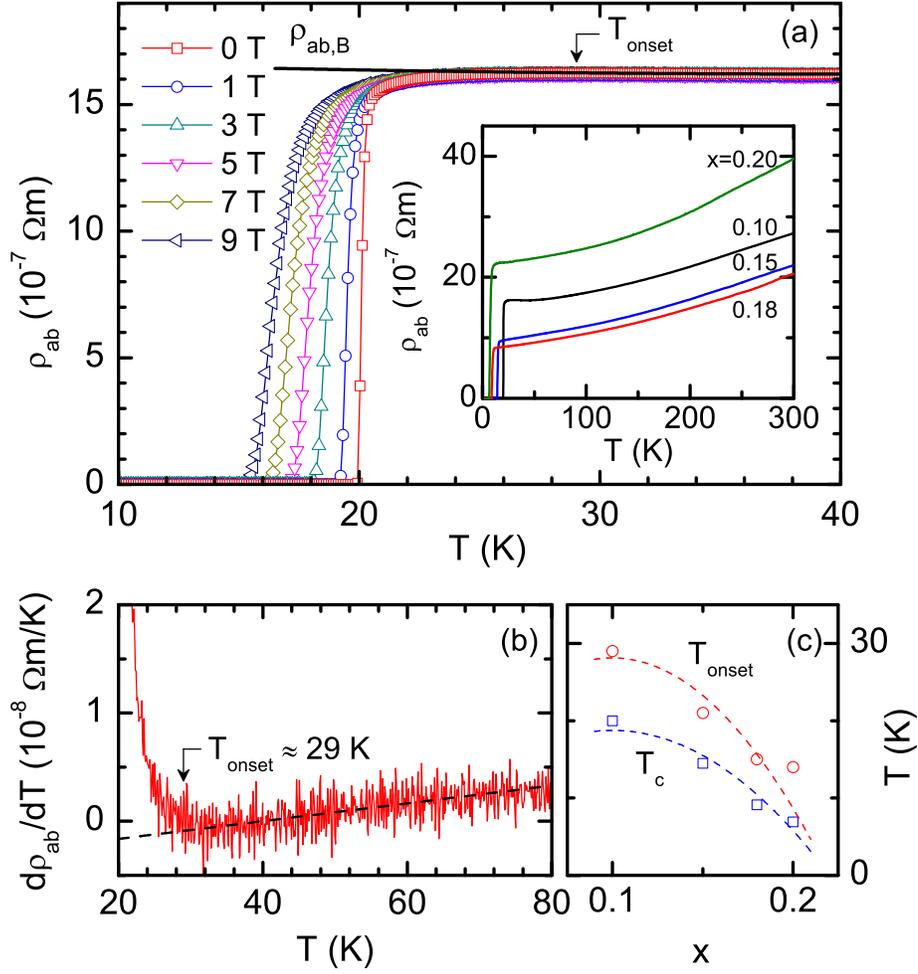}
\caption{(a) Detail of the resistive transition corresponding to the $x=0.1$ crystal. The thick line is the background contribution. Inset: overview up to room temperature for all crystals studied. (b) Temperature dependence of $d\rho_{ab}/dT$ indicating the onset of fluctuation effects for the $x=0.1$ crystal. The line corresponds to the normal-state background. (c) $x$ dependence of $T_c$ and $T_{\rm onset}$. The blue line is a fit of a degree-two polynomial to $T_c(x)$. The red line, obtained as $1.5T_c(x)$, is in relatively good agreement with the observed $T_{\rm onset}$ values. The factor 1.5 is what one would expect from the GGL approach presented in Appendix B by using a cutoff constant of $c=0.4$.}
\label{rho}
\end{center}
\end{figure}

An overview of the $\rho_{ab}(T)$ data up to room temperature is presented in the inset of Fig.~1(a). The seemingly non-monotonous $x$-dependence at 300~K may be explained in terms of the above mentioned geometrical uncertainties. As expected,\cite{growth,olariu} for $x>0.1$, the kinks associated with structural (tetragonal-orthorhombic) and magnetic (paramagnetic-antiferromagnetic) transitions are not observed, and for $x=0.1$ they are irrelevant when compared with the rounding associated to the fluctuations. This is an important experimental advantage in order to determine the superconducting contribution to the electric conductivity, $\Delta \sigma_{ab}=\rho_{ab}^{-1}-\rho_{ab,B}^{-1}$, where $\rho_{ab,B}$ is the normal-state or background contribution. In view of the linear temperature dependence of $d\rho_{ab}/dT$ (an example for $x=0.1$ is presented in Fig.~\ref{rho}(b)), the background contributions were estimated by fitting a quadratic polynomial to the measured $\rho_{ab}(T)_H$ from $\sim4T_c$ down to $T_{\rm onset}$, the temperature below which fluctuation effects are measurable. In turn, this was determined as the temperature at which $d\rho_{ab}/dT$ rises above the extrapolated normal-state behavior beyond the noise level [see Fig. 1(b)]. As shown in Fig.~1(c), $T_{\rm onset}(x)\approx 1.5T_c(x)$ an issue that will be analyzed later. An example of the background contribution (for the $x=0.1$ crystal) is shown in Fig.~1(a). On the scale of the effect of the superconducting fluctuations, $\rho_{ab,B}(T)$ is independent of the applied magnetic field. This further experimental advantage to study the fluctuation-induced in-plane magnetoconductivity is confirmed by the detailed representation of Fig.~2(a), where it is shown that $\sigma_{ab}(0)-\sigma_{ab}(H)$ becomes negligible well above $T_c$. A detailed analysis of the effect on the measured $\Delta\sigma_{ab}$ of the typical uncertainty in the background contribution is presented in Appendix A.

\section{Data analysis}

\subsection{Comparison with the conventional Aslamazov-Larkin approach}

An example (for the $x=0.1$ crystal) of the $\Delta\sigma_{ab}(H)$ dependence at several temperatures just above $T_c$ is presented in Fig.~2(b). In the $H\to 0$ limit $\Delta\sigma_{ab}$ tends to a constant value, in qualitative agreement with the conventional Aslamazov-Larkin (AL) result,\cite{AL}
\begin{equation}
\Delta\sigma_{ab}=\frac{e^2}{32\hbar\xi_c(0)}\varepsilon^{-1/2},
\label{AL}
\end{equation} 
where $e$ is the electron charge, $\hbar$ the reduced Planck constant, and $\xi_c(0)$ the $c$-axis coherence length amplitude. However, $\Delta\sigma_{ab}$ decreases above a temperature-dependent magnetic field which is close to the scale for the observation of finite-field effects (see below).\footnote{It is worth noting that a negative fluctuation magnetoresistance has been observed in granular metallic films in high fields (of the order of the critical field), see Ref.~\cite{negativeMR}. This effect is not observed here even in the vicinity of $T_c$, where superconducting and normal domains may coexist as a consequence of $T_c$ inhomogeneities. On the one side, the magnetic field used in our experiments may not be strong enough as to observe such an effect. On the other, the indirect contributions to the fluctuation conductivity needed to explain the negative magnetoresistance (see Ref.~\cite{negativeMR}) may not be present in non-s wave superconductors, as it may be the case of iron pnictides.} An example of the $\Delta\sigma_{ab}$ dependence on $\varepsilon$ (also corresponding to the $x=0.1$ crystal) is shown in Fig.~3(a). The double-logarithmic scale was chosen to explore in detail the high-$\varepsilon$ region. As may be clearly seen in this figure, the fit of Eq.~(\ref{AL}) to the low-field data (dashed line) is excellent at low reduced temperatures (up to $\varepsilon\approx0.06$), except for $\varepsilon<\Delta T_c/T_c\approx0.015$ where $T_c$ inhomogeneities may play a role (see below). However, a large discrepancy is found at high-$\varepsilon$ values: while the AL theory never vanishes, the experimental $\Delta\sigma_{ab}$ ends up below the experimental resolution for $\varepsilon\sim0.3$ for all the applied magnetic fields. 
A similar rapid falloff of the fluctuation induced conductivity was already observed more than 30 years ago in low-$T_c$ conventional superconductors by Johnson and coworkers,\cite{johnson1,johnson2} who already stressed that this type of behaviour could not be described in terms of a power law in $\varepsilon$. 
Thus, our present data provides further evidence of the failure of those proposals which, like the momentum cutoff approach\cite{shortwavelength} and equivalent microscopic calculations\cite{larkin}
\footnote{As we have already commented in Ref.~\cite{PRBcarballeira}, the microscopic approach proposed in this work results to be equivalent to apply a momentum cutoff in to the GL-theory (see Refs.~\cite{shortwavelength}), since both lead to an asymptotic behaviour of the paraconductivity at high reduced temperatures proportional to epsilon to the -3. Such a behavior was seemingly observed in cuprate superconductors by Varlamov and coworkers (see Fig.~7.6, and the corresponding reference, in that book). Nevertheless, such a result, the only accounted on the paraconductivity in that book, seems to be an artifact of the data analysis and, in any case, it is in contradiction with the measurements published until now by other groups in any low- or high-$T_c$ superconductor (for earlier results see, e.g., Ref.~\cite{shortwavelength} and references therein).}, predict a definite critical exponent for $\Delta\sigma_{ab}$ at high reduced temperatures.

\subsection{Comparison with the GL theory with an energy cutoff in the fluctuations spectrum}

The failure of GL-based approaches to explain the $\Delta\sigma_{ab}$-data at high-$\varepsilon$ has also been observed in HTSC,\cite{PRBcarballeira} and it also appears in the fluctuation diamagnetism of both HTSC \cite{PRLcarballeira,physicaC} and low-$T_c$ alloys.\cite{PRLtinkham,PbIn} It has been attributed to the fact that the GL-theory overestimates the contribution of the high-energy fluctuation modes.\cite{PRLtinkham} 
In fact, although GL approaches are formally valid only in the vicinity of the transition, it was found that its applicability may be extended to the high-$\varepsilon$ region through the introduction of an energy cutoff.\cite{PRBcarballeira,EPLcutoff,PbIn,physicaC} Since the energy of the fluctuation modes increases with $H$, the inclusion of such a cutoff is also needed when analyzing the effect of a finite applied magnetic field on the superconducting fluctuations.\cite{breakdown}
In the case of 3D anisotropic superconductors, the one well adapted to 122 iron arsenides, the paraconductivity in presence of an energy cutoff may be easily calculated by using the procedure proposed in the pioneering work by A.~Schmid.\cite{schmid} These calculations, whose details are presented in Appendix B, lead to
\begin{eqnarray}
&&\Delta\sigma_{ab}=\frac{e^2}{32\hbar \pi\xi_c(0)}\sqrt{\frac{2}{h}}\int_0^{\sqrt{\frac{c-\varepsilon}{2h}}}\mathrm{d}x
\left[\psi^1\left(\frac{\varepsilon+h}{2h}+x^2\right) \right.\nonumber\\
&&\left.-\psi^1\left(\frac{c+h}{2h}+x^2\right)
\right]\,,
\label{magnetocutoff}
\end{eqnarray}
where $\psi^1$ is the first derivative of the digamma function, $h=H/[\phi_0/2\pi\mu_0\xi_{ab}^2(0)]$ is the reduced magnetic field, $\xi_{ab}(0)$ is the in-plane coherence length amplitude, and $c$ is the cutoff constant (expected to be of the order of 0.5)\cite{EPLcutoff}. In the zero magnetic field limit ({\it i.e.}, for $h\ll \epsilon$) and in the absence of cutoff ($c\to\infty$), Eq.~(\ref{magnetocutoff}) reduces to the AL expression, Eq.~(\ref{AL}). It is also worth noting that Eq.~(\ref{magnetocutoff}) leads to the $\Delta\sigma_{ab}$ vanishing at $\varepsilon=c$. In an attempt to check the applicability range of the 3D-anisotropic GL approach under an energy cutoff, in what follows we will compare our experimental data with Eq.~(\ref{magnetocutoff}). 

\subsubsection{Optimally doped sample:} A first check may be carried out through measurements performed with $h=0$, because in this case Eq.~(\ref{magnetocutoff}) depends only on $\xi_c(0)$ and $c$. In the optimally doped sample the fit of Eq.~(\ref{magnetocutoff}) to the $\Delta \sigma_{ab}(\varepsilon,h=0)$ data is excellent above $\varepsilon=0.02$, including the $\Delta\sigma_{ab}$ vanishing at high $\varepsilon$-values [see Fig.~3(a)]. This analysis leads to $\xi_c(0)=0.8$~nm and $c=0.39$. By using these values, Eq.~(\ref{magnetocutoff}) was then fitted to the data obtained with $h>0$ with $\xi_{ab}(0)$ as the only free parameter. The fit quality is also excellent up to the largest field used in the experiments leading to $\xi_{ab}(0)=2.3$~nm. For comparison, the conventional AL approach [Eq.~(\ref{AL})] evaluated with the same $\xi_{ab}(0)$ and $\xi_c(0)$ values is represented as a dotted line in Fig.~3(a).

As expected, the adequacy of Eq.~(\ref{magnetocutoff}) extends to the $\Delta\sigma_{ab}(H)_\varepsilon$ representation of Fig.~2(b), where the solid lines were evaluated by using in Eq.~(\ref{magnetocutoff}) the above $\xi_{ab}(0)$, $\xi_c(0)$ and $c$ values. The dashed line in this figure is the crossover to the region at which finite field effects are expected to be relevant, and was evaluated by using $h=\varepsilon$ in Eq.~(\ref{magnetocutoff}).\footnote{See e.g., Ref.~\cite{PRLtinkham}. This criterion for the presence of finite-field effects may be more intuitively written as $H=\phi_0/2\pi\mu_0\xi_{ab}^2(T)$. This magnetic field scale is sometimes referred to as \textit{ghost critical field}, see e.g., Ref.~\cite{criterio}.}
It is worth mentioning that the in-plane magnetoconductivity, $\sigma_{ab}(0)-\sigma_{ab}(H)$, in Fig.~2(a) is also in excellent agreement with $\Delta\sigma_{ab}(0)-\Delta\sigma_{ab}(H)$ as evaluated from Eq.~(\ref{magnetocutoff}) by using the same $\xi_{ab}(0)$, $\xi_c(0)$ and $c$ values (solid lines). As the $\sigma_{ab}(0)-\sigma_{ab}(H)$ data do not depend on a background subtraction, this fact further validates the procedure used to determine the background contribution to obtain $\Delta\sigma_{ab}$ in Figs.~2(b) and 3.

\subsubsection{Overdoped samples:} An example of the $\Delta\sigma_{ab}(\varepsilon)_H$ dependence corresponding to an overdoped crystal ($x=0.15$) is shown in Fig.~3(b). The upturns observed in the low-$H$ isofields (indicated with arrows) are associated with the above mentioned $T_c$ inhomogeneities inherent to non-optimally-doped samples. It is expected that a magnetic field of the order of $H_{c2}(0)\Delta T_c/T_c\stackrel{>}{_\sim}$~1 T will shift $T_c(H)$ so that $\Delta\sigma_{ab}$ would be unaffected by $T_c$ inhomogeneities down to $\varepsilon=0$. Then, we have fitted Eq.~(\ref{magnetocutoff}) to the experimental data for $\mu_0H\geq3$~T with $\xi_{ab}(0)$, $\xi_c(0)$, and $c$ as free parameters. The agreement is excellent in the entire $\varepsilon$ range, and even extends to the lowest magnetic fields for $\varepsilon$-values well above $\Delta T_c/T_c$. 
A similar result is obtained in the other overdoped crystals studied. The $x$-dependence of the resulting $\xi_{ab}(0)$, $\xi_c(0)$ and $c$ will be analyzed in the next Section.

We have checked that the $T_c$-inhomogeneities model proposed in Ref.~\cite{maza}, which is based on Bruggeman's effective medium theory,\cite{bruggeman} accounts for the upturn in the $H=0$ measurement. According to this model, the temperature dependence of effective electrical conductivity in presence of $T_c$ inhomogeneities, $\sigma_{ab}^{\rm eff}(T)$, is obtained by numerically solving
\begin{equation}
\int\frac{\sigma_{ab}(T,T_c)-\sigma_{ab}^{\rm eff}(T)}{\sigma_{ab}(T,T_c)+2\sigma_{ab}^{\rm eff}(T)}\delta(T_c){\rm d}T_c=0,
\end{equation}
where 
$\delta(T_c)$ is the distribution of critical temperatures in the sample, and $\sigma_{ab}(T,T_c)=\rho_{ab,B}^{-1}(T)+\Delta \sigma_{ab}(T,T_c)$ with $\Delta\sigma_{ab}(T,T_c)$ given by Eq.~(\ref{magnetocutoff}). The dot-dashed line shown in Fig.~3(b) was obtained by just assuming for $\delta(T_c)$ a Gaussian distribution 1.2 K wide (FWHM), which is in reasonable agreement with the $\Delta T_c$ value determined for this sample. 

\section{Discussion of the results}

\subsection{Dependence of the superconducting parameters on the doping level}

The $x$-dependence of the coherence lengths resulting from the above analysis is presented in the inset in Fig.~3(a). $\xi_{ab}(0)$ grows monotonically with $x$, whereas $\xi_c(0)$ is roughly independent of $x$. As a consequence, the anisotropy factor $\gamma=\xi_{ab}(0)/\xi_c(0)$ grows with $x$ from $\sim 2.9$ in the optimally doped crystal, up to above $\sim10$ in the most overdoped crystal ($x=0.2$). Such a large $\gamma$ value is, however, consistent with the 3D behavior of this material because the corresponding $2\xi_c(0)$ is still larger than the Fe-layers separation. Previous studies in BaFe$_{2-x}$Ni$_x$As$_2$ focus on samples close to optimal doping, and report superconducting parameters consistent with our present results.\cite{tao,sun09,ni10,shahbazi11} Although the large $\gamma$ value observed in the highly overdoped BaFe$_{1.8}$Ni$_{0.2}$As$_2$ was never observed in a 122 compound, some studies in the very similar BaFe$_{2-x}$Co$_x$As$_2$ reported that $\gamma$ increases on increasing $x$ slightly above the optimal value for this compound.\cite{ni10,ni08,vinod} This effect was associated with the change in the structural/magnetic state of the system,\cite{ni10,ni08} or with an increase in the ratio of inter/intraband coupling.\cite{vinod} 

\subsection{Temperature onset for the fluctuation effects}

Given the relation $c=\ln(T_{\rm onset}/T_c)$, the cutoff constant ($c\approx0.3-0.7$) turned out to be consistent with the $T_{\rm onset}$ values in Fig.~1, roughly 1.5$T_c(x)$. In a recent study of other iron pnictides (Ref.~\cite{kumar}) a change is observed in the Raman modes below a comparable temperature ($\sim2 T_c$) which is also attributed to superconducting fluctuations. Remarkably, such a $T_{\rm onset}/T_c$ ratio is also within the values observed in other superconducting families, including conventional low-$T_c$ superconductors\cite{PbIn,breakdown}, MgB$_2$ \cite{MgB2}, NbSe$_2$ \cite{NbSe2}, and HTSC not severely affected by $T_c$ inhomogeneities.\cite{physicaC,HTSC}
\footnote{Note that some works report the observation of $\varepsilon_{\rm onset}$-values different from 0.5 in non-optimally-doped HTSC and amorphous low-$T_c$ superconductors. See e.g., Ref.~\cite{HTSC2}. However, these differences in the $\varepsilon_{\rm onset}$ values could be related to the strong $T_c$ dependence on the stoichiometry in these materials, and the unavoidable presence of stoichiometric inhomogeneities, see Ref.~\cite{intrinsic} and \cite{commentrourke}.} This study extends to Fe-based superconductors the proposal that superconducting fluctuations vanish at the temperature at which the superconducting wavefuntion shrinks to lengths of the order of the pairs size.\cite{EPLcutoff}

\subsection{Relevance of phase fluctuations or indirect contributions to $\Delta\sigma$}

A direct consequence of the excellent agreement of the GL theory to explain our data is the absence of appreciable indirect contributions to the fluctuation-induced in-plane conductivity and magnetoconductivity of Fe-based  superconductors, including the \textit{Maki-Thompson} (MT) and the so-called \textit{density of states} (DOS) contributions. These indirect contributions have been found to be negligible also in HTSC,\cite{pomar} and in both systems this could be attributed to the non $s$-wave pairing.\cite{yip} 

Finally, it has recently been claimed that conventional GL approaches for the fluctuation effects above $T_c$ are not applicable at low field amplitudes (typically below 1~T) in SmFeAsO$_{0.8}$F$_{0.2}$, due to the presence of \textit{phase fluctuations}.\cite{prando} The relevance of phase fluctuations has also been proposed for a member of the less anisotropic 122 family, Ba$_{1-x}$K$_x$Fe$_2$As$_2$.\cite{sugui} However, our present results show that the anomalous increase of fluctuation effects at low field amplitudes may be explained in the framework of GL approaches by taking into account the effect of $T_c$ inhomogeneities. This could suggest that the effect of phase fluctuations in these materials is less important than previously estimated.

\section{Conclusions}

We have presented detailed measurements of the fluctuation-induced in-plane conductivity and magnetoconductivity in a series of high quality BaFe$_{2-x}$Ni$_x$As$_2$ single crystals covering doping levels from the optimal one ($x=0.1$) up to the highly overdoped ($x=0.2$). The sharp superconducting transition allowed us to investigate fluctuation effects in almost all the accessible reduced-temperature window above $T_c$. In turn, the smooth temperature and field dependences of the normal-state in-plane resistivity permitted a reliable estimation of the temperature onset for the fluctuation effects. 
As an example of the usefulness of these data to probe the different approaches for the superconducting fluctuations around $T_c$ in iron pnictides, we have also presented here a detailed comparison with a version of the phenomenological Gaussian Ginzburg-Landau approach that includes an energy cutoff, which reduces to the well-known Aslamazov-Larking result at low reduced-temperatures and magnetic fields. Our data are in good agreement with this approach in all the studied temperature and field regions, suggesting the absence, even in the very low magnetic field regime, of local superconducting order associated with phase fluctuations,\cite{sugui,prando} at present a debated aspect of the HTSC.\cite{parker,bilbro,kondo,rourke,ramallo12}  Other contributions to the fluctuation-induced in-plane magnetoconductivity (Maki-Thompson, Zeeman, and DOS) are found to be negligible, as is also the case in HTSC.\cite{pomar}
Finally, we have also shown that when analyzing the low-field regime, it is crucial to take into account the $T_c$-inhomogeneities associated with chemical disorder, which are always present to some extent even in the best crystals, due to the unavoidable random distribution of doping ions.\cite{intrinsic} It would be interesting to extend our present analysis to the critical region by using the LLL scaling proposed in Ref.~\cite{tesanovic}, and also check our present results in other families of Fe-based superconductors, in particular in the more anisotropic 1111 pnictides, where the fluctuations dimensionality is at present a debated issue.\cite{flucC,pallecchi,fanfarillo,kim,putti,pandya}

\ack

Supported by the Spanish MICINN and ERDF \mbox{(No.~FIS2010-19807)}, and by the Xunta de Galicia (Nos.~2010/XA043 and 10TMT206012PR). SSS and ADA acknowledge support from the CNPq. The work at IOP, CAS in China is supported by NSFC Program (No.~11004233) and 973 project (No.~2011CBA00110).

\vspace{1cm}

\textit{Note added in proof:} After this paper had been submitted, we became aware of a work about superconducting fluctuations in transport properties of LiFeAs \cite{noteadded}. In contrast with our present results, and in spite of the three-dimensional character of this compound, it is claimed that the fluctuation conductivity agrees with the two-dimensional Aslamazov-Larkin expression, and also that the onset temperature is as high as $\sim2.5T_c$.

%
%

%
%
\begin{figure}[b]
\begin{center}
\includegraphics[scale=.4]{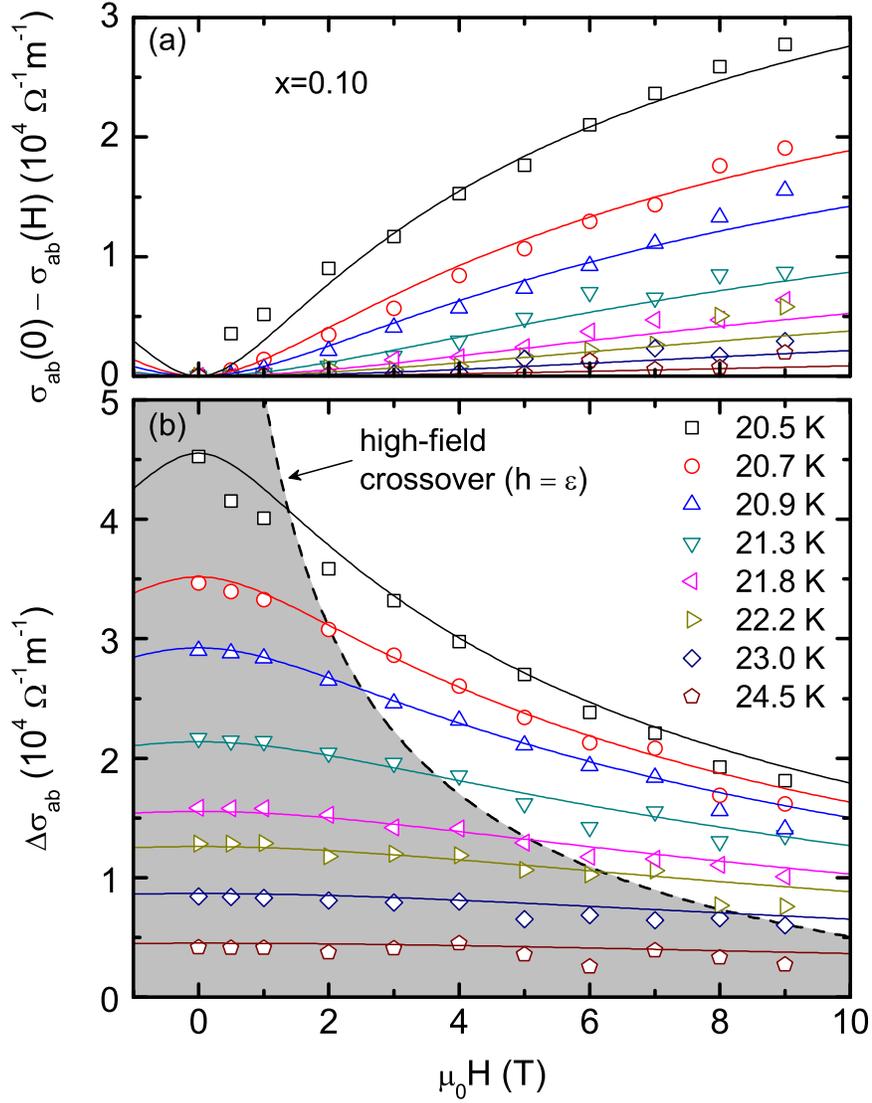}
\caption{Example (for the $x=0.1$ crystal) of the $H$ dependence of the fluctuation-induced in-plane magnetoconductivity (a), and conductivity (b), for temperatures just above $T_c$. The lines correspond to $\Delta\sigma_{ab}(0)-\Delta\sigma_{ab}(H)$ and $\Delta\sigma_{ab}(H)$ respectively, evaluated by using Eq.~(\ref{magnetocutoff}) and the same $\xi_{ab}(0)$, $\xi_c(0)$ and $c$ values. The coincidence between $\Delta\sigma_{ab}(0)-\Delta\sigma_{ab}(H)$ and $\sigma_{ab}(0)-\sigma_{ab}(H)$ (which is background-independent), represents an important verification of the adequacy of the procedure used to subtract the background contribution. The dashed line in (b), evaluated by using $h=\varepsilon$ in Eq.~(\ref{magnetocutoff}), represents the crossover to the region at which finite field effects are important (shadowed).\cite{criterio}}
\label{vse}
\end{center}
\end{figure}

%
%
\begin{figure}[t]
\begin{center}
\includegraphics[scale=.7]{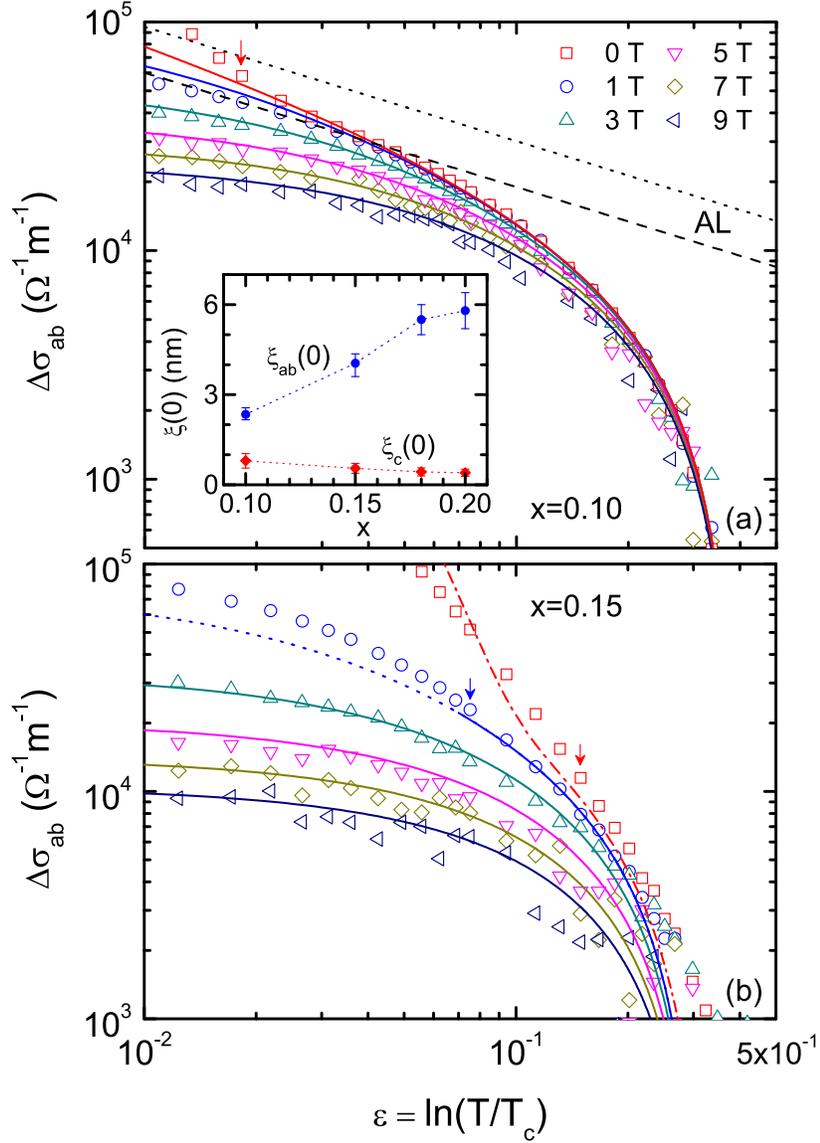}
\caption{Examples of the $\Delta\sigma_{ab}$ dependence on $\varepsilon$ for the crystals with $x=0.1$ (a) and 0.15 (b). The conventional AL approach [dashed line in (a)] fits the data in the low-$\varepsilon$ and -$H$ region but is not valid above $\varepsilon\sim0.1$. In contrast, Eq.~(\ref{magnetocutoff}) (solid lines) reproduces the $\Delta\sigma_{ab}$ reduction at high-$\varepsilon$ and -$H$ values. The $\xi_{ab}(0)$ and $\xi_c(0)$ values resulting from the fits of Eq.~(\ref{magnetocutoff}) for all doping levels are presented in the inset, where the error bars include the uncertainty in the $\rho_{ab}$ amplitude. For comparison, the AL approach evaluated with the same parameters is represented as a dotted line in (a). The upturns at low-$H$ (indicated by arrows) are due to $T_c$ inhomogeneities associated with the random distribution of Ni dopants. The dot-dashed line in (b) corresponds to the effective-medium approach proposed in Ref.~\cite{maza} to account for the effect of $T_c$ inhomogeneities, evaluated by assuming a Gaussian $T_c$ distribution 1.2 K wide.}
\label{vsh}
\end{center}
\end{figure}

\clearpage

\appendix

\section{Uncertainty in $\Delta\sigma_{ab}(\varepsilon,H)$ associated with the uncertainties in the transition temperature and in the normal-state contribution.}

Here we present some representative examples of how the uncertainty in the transition temperature and in the normal state contribution affect the determination of the fluctuation contribution to the in-plane electrical conductivity.

\begin{figure}[b]
\begin{center}
\includegraphics[scale=.8]{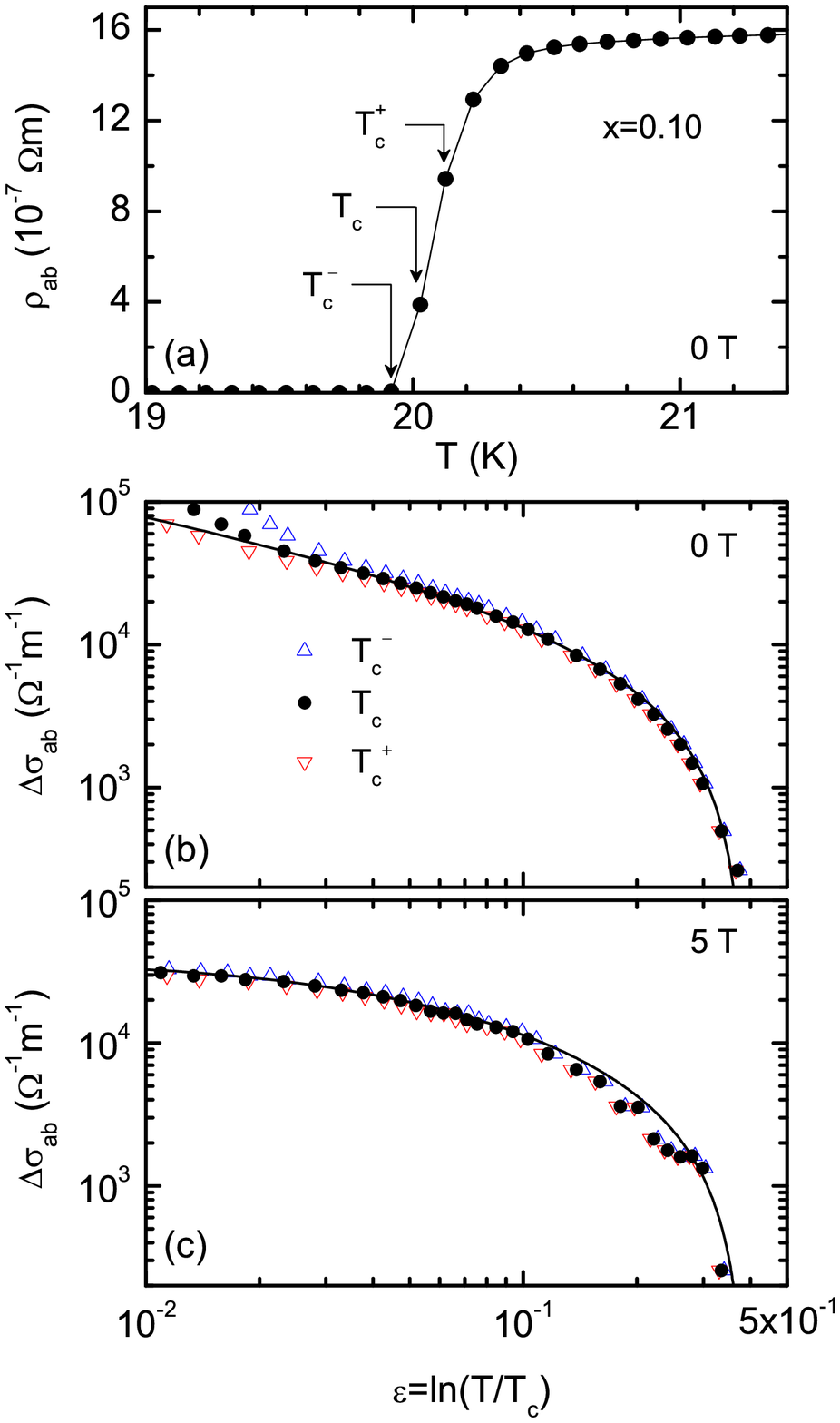}
\caption{(a) Detail of the resistive transition corresponding to the $x=0.1$ crystal, indicating the upper, lower, and midpoint $T_c$ values. The effects of the $T_c$ choice on the resulting $\Delta\sigma_{ab}(\varepsilon)$ is presented in (b) and (c) for $\mu_0H=0$ and 5 T, respectively. The lines are fits of Eq.~(\ref{magnetocutoff}) to the solid data points.}
\label{fig4}
\end{center}
\end{figure}

In our work $T_c$ is estimated as the temperature at which the slope of the $\rho_{ab}(T)$ curve is maximum. In turn, the $T_c$ uncertainty was estimated as $\Delta T_c=2(T_c-T_c^-)$, where $T_c^-$ is the temperature at which $\rho_{ab}$ vanishes. In Fig.~\ref{fig4}(a) we present a detail of the zero-field resistive transition of the optimally-doped crystal (with $x=0.1$) showing the locations of $T_c$, $T_c^-$, and $T_c^+=T_c+\Delta T_c/2$ (the latter represents an upper limit for the $T_c$ value). In Fig.~\ref{fig4}(b) we show the $\Delta\sigma_{ab}$ dependence on the reduced temperature, $\varepsilon=\ln(T/T_c)$, as obtained by using those three representative $T_c$ values. As may be clearly seen, $\Delta\sigma_{ab}$ is almost unaffected by the uncertainty in $T_c$ above $\varepsilon=0.03$. The line is the best fit of Eq.~(\ref{magnetocutoff}) above this $\varepsilon$ value to the solid data points. In the presence of a finite magnetic field $\Delta\sigma_{ab}$ presents a smoother behavior at $T_c$ (its divergence is shifted to lower temperatures) and $\Delta\sigma_{ab}(\varepsilon)$ is less affected by the uncertainty in $T_c$. This is illustrated in Fig.~\ref{fig4}(c), where the result corresponding to $\mu_0H= 5$~T is presented. The solid line in this figure is the best fit of Eq.~(\ref{magnetocutoff}) to the solid data points.

\begin{figure}[b]
\begin{center}
\includegraphics[scale=.8]{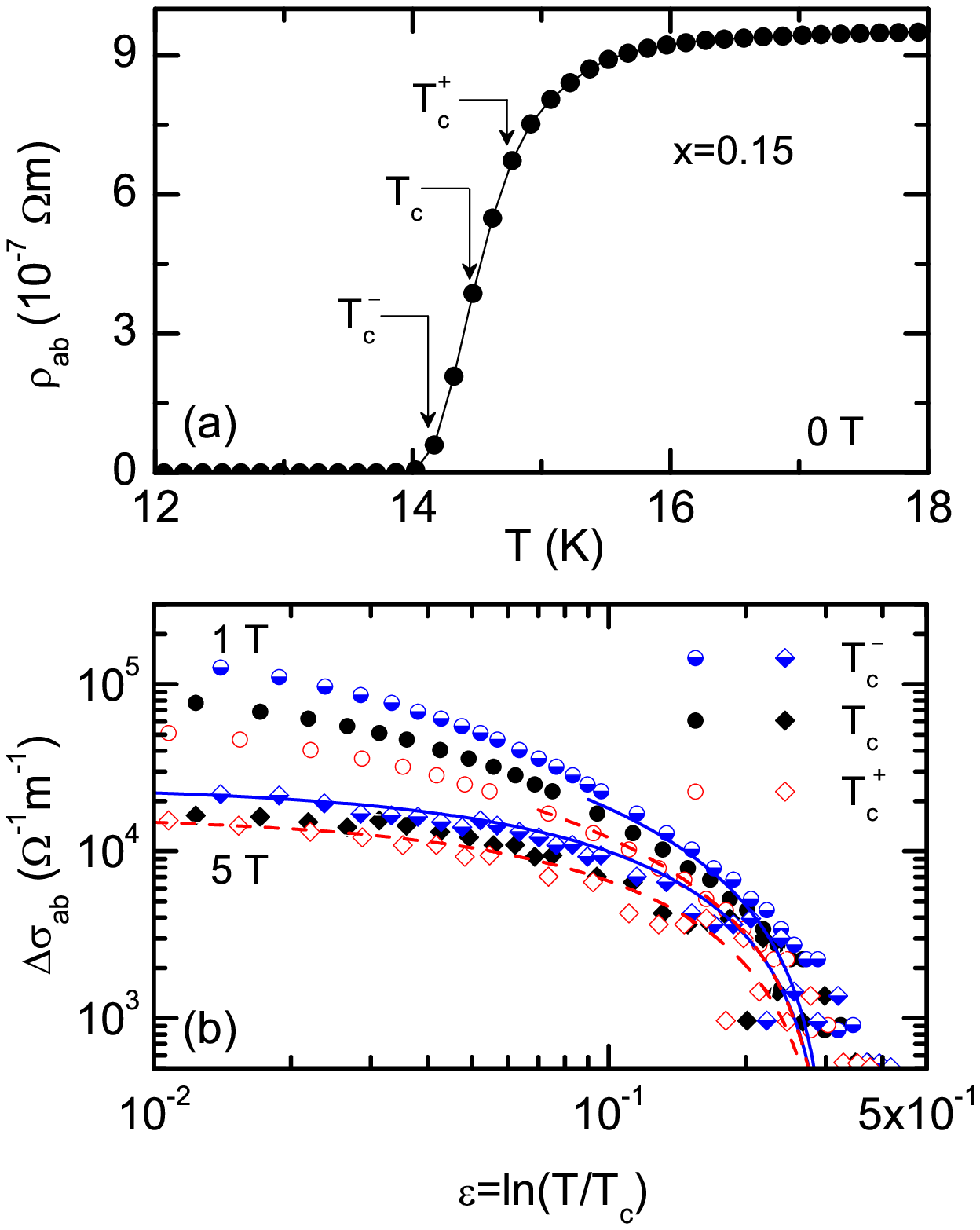}
\caption{(a) Detail of the resistive transition corresponding to the $x=0.15$ crystal, indicating the upper, lower, and midpoint $T_c$ values. The effects of the $T_c$ choice on the resulting $\Delta\sigma_{ab}(\varepsilon)$ is presented in (b) for $\mu_0H=1$ and 5 T. The solid (dashed) lines are fits of Eq.~(\ref{magnetocutoff}) to the data obtained with $T_c^-$ ($T_c^+$). The difference in the resulting $\xi_c(0)$ is about $\pm20\%$. }
\label{fig5}
\end{center}
\end{figure}

The same analysis is presented in Fig.~\ref{fig5} for one of the overdoped samples ($x=0.15$). In this case, as may be seen in the detail of the resistive transition presented in Fig.~\ref{fig5}(a), $\Delta T_c/T_c$ is larger than in the optimally-doped crystal.  As commented above, this may be attributed to the random distribution of Ni dopants and to the $T_c$ dependence on the doping level. This effect is also present in doped cuprates (see, e.g., Ref. \cite{intrinsic}) and may be intrinsic to non-optimally-doped superconductors. The $\Delta\sigma_{ab}$ dependence on the reduced temperature, as obtained by using the three representative $T_c$ values shown in Fig.~\ref{fig5}(a), is presented in Fig.~\ref{fig5}(b). Circles and rhombus were obtained with  $\mu_0H=1$~T and 5 T, respectively. The effect on $\Delta\sigma_{ab}$ of changing $T_c$ by $T_c^-$ or $T_c^+$  may still be accounted for by Eq.~(\ref{magnetocutoff}) by just changing the $\xi_c(0)$ value (which affects the $\Delta\sigma_{ab}$ amplitude) by $\pm$20\%.

\begin{figure}[t]
\begin{center}
\includegraphics[scale=.7]{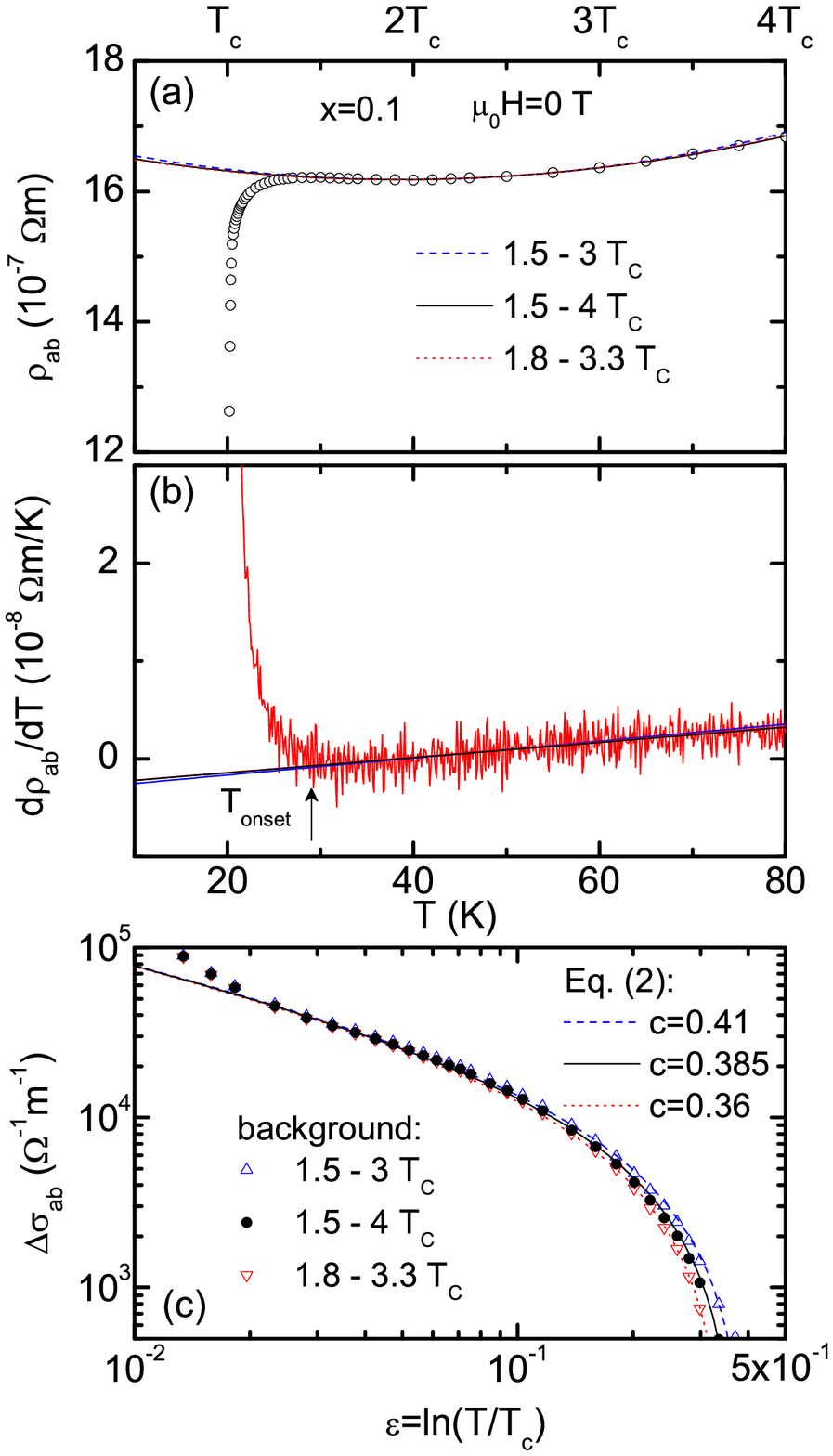}
\caption{(a) Detail of the normal-state in-plane resistivity for the $x=0.1$ crystal, showing the robustness of the background contribution to changes in the temperature interval used to determine it. The same backgrounds are presented in (b) in the $d\rho_{ab}/dT$ vs. $T$ representation. The effects of the background choice on the resulting $\Delta\sigma_{ab}(\varepsilon)$ is presented in (c), where the lines are fits of Eq.~(\ref{magnetocutoff}) to the different data sets.}
\label{fig6}
\end{center}
\end{figure}

Next, we present an example (corresponding to the $x=0.1$ crystal) of the typical uncertainty in $\Delta\sigma_{ab}$ associated with the determination of the background contribution. In Fig.~\ref{fig6}(a) we present a detail of the temperature dependence of the in-plane resistivity in the normal state up to $\sim4T_c$. The lines are the background contributions as determined by fitting a degree-two polynomial in different temperature intervals between the onset temperature for the fluctuation effects ($T_{\rm onset}$) and $4T_c$. The robustness of the background contribution to changes in the fitting region is a consequence of the smooth behavior of the normal-state in-plane resistivity in a wide temperature region above $T_{\rm onset}$ (it varies about 1\% from $T_c$ up to $\sim1.5T_c$). The same background contributions are presented in the $d\rho_{ab}/dT$ vs. $T$ representation in Fig.~\ref{fig6}(b). The well defined linear behavior of $d\rho_{ab}/dT$ up to $\sim 4T_c$ justifies the use of a quadratic form to determine $\rho_{ab,B}(T)$. This last figure also illustrates that $T_{\rm onset}$ (estimated as the temperature at which $d\rho_{ab}/dT$ rises above the extrapolated normal-state behavior beyond the noise level) is almost independent of changes in the background fitting region. In Fig.~\ref{fig6}(c) we present the $\varepsilon$-dependence of $\Delta\sigma_{ab}$ as obtained by using the background contributions in Fig.~\ref{fig6}(a). Solid lines correspond to Eq.~(\ref{magnetocutoff}) evaluated with $\xi_c(0)=0.8$~nm and the values indicated for the cutoff constant. As is clearly shown, the uncertainty in the background leads to an uncertainty in the cutoff constant below $\pm10$\%. Given the relation $c=\ln(T_{\rm onset}/T_c)$, this leads to an uncertainty in $T_{\rm onset}$ which remains below $\pm4$\%.

%
%

%
%

%
%

\clearpage

\section{3D anisotropic GGL approach for $\Delta\sigma_{ab}$ in the finite-field regime}

Our starting point is the model proposed by A.~Schmid (adapted for a 3D anisotropic superconductor), based on
a combination of the standard GL-expression for the thermally-averaged
current density of the superconducting condensate,
\beq
{\bf J}=\frac{\hbar}{i} \frac{e}{m^{*}}\langle\Psi^{*}\nabla\Psi-
\Psi\nabla\Psi^{*}\rangle-\frac{4e^{2}}{m^{*}c} {\bf A}(t)
\langle\Psi^{*}\Psi\rangle,
\label{corriente}
\eeq
with the generalized Langevin equation of the order parameter, \cite{schmid}
\begin{eqnarray}
&&-\frac{\pi \hbar a_{0}}{8k_{B}T_{c0}}\left(
\frac{\partial}{\partial t}+
\frac{2ie}{\hbar c}\Phi({\bf r})
\right)
\Psi({\bf r}, t)=
\nonumber
\\
&&\left[
a_{0}\epsilon+\frac{1}{2m^{*}}
\left(\frac{\hbar}{i}\nabla-\frac{2e}{c}{\bf A}( t)
\right)^{2}
\right]
\Psi({\bf r}, t)
+G({\bf r}, t).
\label{tdgl}
\end{eqnarray}
In these equations, $m^{*}$ is the
mass of the Cooper pairs, $c$ is the speed of light, $\Phi$ is
the electrical potential, $k_{B}$ is the Boltzmann constant,
$a_{0}=\hbar^{2}/2m^{*}\xi^{2}(0)$ is the Ginzburg-Landau normalization
constant [here $\xi(0)$ is the GL coherence length amplitude], and ${\bf  A}$ is the vector potential. Note also that Eq.~(\ref{tdgl}) almost coincides with the conventional
time-dependent Ginzburg-Landau equation of the order parameter.
The only difference is the presence of
a random force, $G({\bf r},t)$, which must be completely
uncorrelated
in space and time. The latter implies that  $G({\bf r},t)$ will
verify
\beq
\langle G^{*}({\bf r},t) G({\bf r}',t')\rangle=a\delta({\bf r}-{\bf r}')
\delta(t-t'),
\label{langevin}
\eeq
where $a$ is a normalization constant that may be determined by
just taking into account that in the stationary limit
Eq.~(\ref{tdgl}) has to reproduce the equilibrium thermal average
of the squared order parameter.\cite{schmid} This directly leads to
$a=\pi\hbar a_{0}/4$.

The presence of an homogeneous electrical field, {\bf E}, applied at the
instant $t = 0$ may be taken into account in this formalism
by applying
\beq
{\bf A}(t) =\left\{\begin{array}{ll}
-c {\bf E} t & \mbox{if} \; t \geq 0\\
0 & \mbox{if} \; t<0
\end{array}
\right.
\eeq
and $\Phi({\bf r})=0$ to Eqs.~(\ref{corriente}) and (\ref{tdgl}). After a standard Fourier-like expansion of
the order parameter this leads to
\beq
{\bf J}=V^{-1}\sum_{{\bf p}}\frac{2e}{m^{*}}
(\hbar {\bf p}+2e{\bf E}t)\langle|\Psi_{{\bf p}}|^{2}\rangle,
\label{kcorriente}
\eeq
and, respectively,
\begin{eqnarray}
&&\frac{\partial \Psi_{{\bf p}}(t)}{\partial t}=\nonumber\\
&&-\frac{8k_{B}T_{c0}}{\pi \hbar a_{0}}
\left[a_{0}\epsilon+\frac{(\hbar {\bf p}+2e{\bf E}t)^{2}}{2m^{*}}
\right]\Psi_{{\bf p}}(t)
+G_{{\bf p}}(t),
\label{ktdgl}
\end{eqnarray}
where $\Psi_{{\bf p}}(t)$ is the Fourier-component of the order
parameter corresponding to the wavevector
${\bf p}\equiv (p_{x},p_{y},p_{z})$ and $G_{{\bf p}}(t)$ represents the random
force in momentum space that, accordingly to Eq.~(\ref{langevin}),  will verify
\beq
\langle G^{*}_{{\bf p}}(t) G_{{\bf p}'}(t')\rangle=a\delta({\bf p}-{\bf p}')
\delta(t-t').
\label{klangevin}
\eeq
The thermally averaged current density at an arbitrarily high
electrical field may be now obtained by introducing in
Eq.~(\ref{kcorriente}) the $\langle |\Psi_{{\bf p}}|^{2}\rangle$-expression
resulting from Eq.~(\ref{ktdgl}). The latter is a differential equation with solution
\begin{eqnarray}
&&\Psi_{{\bf p}}(t)=\frac{8k_{B}T_{c0}}{\pi \hbar a_{0}}
\int_{-\infty}^{t} {\rm d}t' G_{{\bf p}}(t)\times
\nonumber
\\
&&\exp\left\{
-\frac{8k_{B}T_{c0}}{\pi \hbar a_{0}}
\int_{t'}^{t} {\rm d}t''
\left[
a_{0}\epsilon+\frac{(\hbar {\bf p}+2e{\bf E}t'')^{2}}{2m^{*}}
\right]
\right\}
\label{soldif}.
\end{eqnarray}
Then, by using Eqs.~(\ref{klangevin}) and ~(\ref{soldif}),
we obtain
\begin{eqnarray}
&&\langle |\Psi_{{\bf p}}|^{2}\rangle=
\frac{16k_{B}^{2}T_{c0}^{2}}{\pi \hbar a_{0}}
\int_{-\infty}^{t} {\rm d}t'
\times
\nonumber
\\
&&\exp\left\{
-\frac{16k_{B}T_{c0}}{\pi \hbar a_{0}}
\int_{t'}^{t} {\rm d}t''
\left[
a_{0}\epsilon+\frac{(\hbar {\bf p}+2e{\bf E}t'')^{2}}{2m^{*}}
\right]
\right\},
\label{squared}
\end{eqnarray}
and, subsequently, the supercurrent density at high applied
electrical fields will be given by
\begin{eqnarray}
&&{\bf J}=
\frac{32ek_{B}^{2}T_{c0}^{2}}{\pi \hbar m^{*}a_{0}V}
\sum_{{\bf p}}
(\hbar {\bf p}+2e{\bf E}t)
\int_{-\infty}^{t} {\rm d}t' \times
\nonumber
\\
&&\exp\left\{
-\frac{16k_{B}T_{c0}}{\pi \hbar a_{0}}
\int_{t'}^{t} {\rm d}t''
\left[
a_{0}\epsilon+\frac{(\hbar {\bf p}+2e{\bf E}t'')^{2}}{2m^{*}}
\right]
\right\}.
\label{sumcurrent}
\end{eqnarray}
As addressed in Ref.~\cite{schmid}, to solve the time-integrations
involved in Eq.~(\ref{sumcurrent}) it is convenient to introduce first
the factor $(\hbar {\bf p}+2e{\bf E}t)$ inside the integral over $t'$ and, then,
to apply the following changes of variables
\beq
\begin{array}{l}
\hbar{\bf p}= \hbar{\bf k}-e{\bf E}(t+t') \\
t'=u+t\\
t''=\frac{1}{2}(u-u')+t\; \; .
\end{array}
\eeq
Equation~(\ref{sumcurrent}) is then transformed into
\begin{eqnarray}
&&{\bf J}=\frac{32ek_{B}^{2}T_{c0}^{2}}{\pi \hbar m^{*}a_{0}V}
\sum_{{\bf k}}
\int_{-\infty}^{0} {\rm d}u
(\hbar {\bf k}-e{\bf E}u)\times
\nonumber
\\
&&\exp\left\{
\frac{8k_{B}T_{c0}}{\pi \hbar a_{0}}
\int_{-u}^{u} {\rm d}u'
\left[
a_{0}\epsilon
+\frac{(\hbar {\bf k}-e{\bf E}u')^{2}}{2m^{*}}
\right]
\right\},
\label{newsumcurrent}
\end{eqnarray}
an expression that solving the integration over $u'$ and introducing the dimensionless variable
$x=\frac{16k_{B}T}{\pi\hbar}\epsilon u$ reads as
\begin{eqnarray}
&&{\bf J}=-
\frac{e^2 \xi^2(0) \pi \bf {E}}{4 \hbar \epsilon^2V}
\sum_{{\bf k}}
\int_{-\infty}^{0} {\rm d}x \,
x\times\nonumber
\\
&&\exp\left[
x+\xi^2(\epsilon){\bf k}^2 x+\left(\frac{E}{E^{*}}
\right)^{2}x^3
\right],
\label{currentx}
\end{eqnarray}
where $E^{*}=\frac{16 \sqrt{3}}{\pi}\frac{k_{B}T_{c}}{e \xi(0)}
\epsilon^{3/2}$ is a temperature-dependent electrical field characteristic of each material that
governs the non-Ohmic regime of the fluctuation induced supercurrent that appears for $E\stackrel{>}{_\sim} E^{*}$. Note also that in this last expression
we have already omitted de $\hbar {\bf k}$-term that appears in the integral over
$u$ in Eq.~(\ref{newsumcurrent}). The reason is that such term is odd and, thus,
the contributions of the addends corresponding to ${\bf k}$ and $-{\bf k}$ cancel each other when carrying out the sum over
the momentum spectrum.

In what follows we will restrict ourselves to applied electrical fields
verifying $E\ll E^{*}$. The last term in the exponential
of Eq.~(\ref{currentx}) can be then suppressed, and the current density exhibits the linear dependence with the electrical
field characteristic of the Ohmic-regime. By using ${\bf J}=\Delta\sigma {\bf E}$ and performing
the trivial integral over $x$, we find the following expression for the fluctuation-induced conductivity as
sum over the modes of the spectrum of the fluctuations
\begin{eqnarray}
\Delta \sigma=\frac{e^2 \xi^2(0) \pi }{4 \hbar \epsilon^2V}
\sum_{{\bf k}}
\frac{1}{[1+\xi^2(\epsilon){\bf k}^2]^{2}}\, \,,
\label{currentx_ohmic_modes}
\end{eqnarray}
that transforming the $k$-sumations into $k$-integrals through
\beq
\sum_{{\bf k}} \to V \int
\frac{{\rm d}k_{x}}{2\pi} \int
\frac{{\rm d}k_{y}}{2\pi}
\int  \frac{{\rm d}k_{z}}{2\pi} \, \, ,
\label{sumtoint}
\eeq
may be rewritten as
\begin{eqnarray}
\Delta \sigma=
\frac{e^2 \xi^2(0)  }{32 \pi^2 \hbar }
\int
 \int
\int
\frac{{\rm d}k_{x}{\rm d}k_{y}{\rm d}k_{z}}{[\epsilon+\xi^2(0){\bf k}^2]^2}\, \,.
\label{currentx_ohmic_integral}
\end{eqnarray}

Equation~(\ref{currentx_ohmic_integral}) has been derived in the framework of a
general model for a $3D$ isotropic superconductor. However, in view of the ratio between the superconducting coherence length amplitudes in BaFe$_{2-x}$Ni$_x$As$_2$ superconductors, the $3D$ anisotropic scenario seems to be more appropriate for these compounds. In this last dimensional case, the scale variation of the order parameter in each spatial direction is determined by the corresponding superconducting coherence length. Thus, considering that the $x$ and $y$ directions lay on the $ab$-plane and that $z$ corresponds to the crystallographic $c$-axis, Eq.~(\ref{currentx_ohmic_integral}) can be adapted to anisotropic $3D$ superconductors by applying $\xi^2(0){\bf k}^2 \rightarrow\xi_{ab}^2(0)k_{x}^2+\xi_{ab}^2(0)k_{y}^2+\xi_{c}^2(0)k_{z}^2$
[here $\xi_{ab}(0)$ is the in-plane superconducting coherence length amplitude]. Using polar coordinates for the $xy$-plane this leads to\footnote{Note that this equation is different from the one that may be obtained by means of the Kubo formula [see, {\it e.g.}, Ref.~\cite{nota}. Both integral cores show essentially a $k_{xy}^{-3}$-dependence, but achieved with different $k_{xy}$-powers in the numerator and denominator. Thus, at zero applied magnetic field and without limits on the $k$-integrals both methods lead to the same result [Eq.~(\ref{AL})]. However for $H \neq 0$ or considering a cutoff in the energy spectrum the final $\Delta \sigma$-expressions show slight quantative (but not qualitative) differences that deserve further studies [{\it cf.} Ref.~\cite{nota} and, respectively, Ref.~\cite{PRBcarballeira}.]}
\begin{eqnarray}
\Delta \sigma_{ab}=
\frac{e^2 \xi_{ab}^2(0) }{16 \pi\hbar  }
\int
\int
\frac{k_{xy}{\rm d}k_{xy}{\rm d}k_{z}}{[\epsilon+\xi_{ab}^2(0)k_{xy}^2+\xi_{c}^2(0)k_{z}^2]^2}.
\label{sigmamomentos}
\end{eqnarray}

If an external magnetic field $H$ is applied parallel to the $c-$direction, the in-plane
spectrum of the fluctuations becomes equivalent to that
of a charged particle in a magnetic field.\cite{landau3} Thus, $k_{xy}$ in Eq.~(\ref{sigmamomentos})
must be replaced by $ \frac{4 e \mu_{0}H}{\hbar}\left(n+\frac{1}{2}\right)$, where $\mu_{0}$ is the vacuum magnetic permeability and $n=0,1\dots$ is the Landau-level index.
As a consequence, the integral
with respect to $k_{xy}$ is transformed into a sum over $n$ through
$\frac {1}{2\pi}\int k_{xy}{\rm d}k_{xy}\rightarrow \sum_{n}$. Besides, we must also include as a
multiplier to Eq.~(\ref{sigmamomentos}) the so-called Landau degeneracy factor
given by $\frac{e\mu_{0}H}{\pi \hbar} = \frac{\mu_{0}H}{\phi_{0}}$ (here $\phi_{0}$ is the magnetic flux quantum).
The resulting expression for $\Delta \sigma_{ab}$ is
\begin{equation}
\Delta \sigma_{ab}=
\frac{e^2h}{16 \pi \hbar }
\int{\rm d}k_{z}
\sum_n [\epsilon+h(2n+1)+\xi_{c}^2(0)k_{z}^2]^{-2},
\label{summagneto}
\end{equation}
where $h=H/H_{c2}(0)$ is the reduced magnetic field and $H_{c2}(0)=\phi_0/2\pi\mu_0\xi_{ab}^2(0)$ is the
upper critical magnetic field perpendicular to the $ab$-planes, linearly extrapolated to $T=0 \; K$.

To take into account the limits imposed by the uncertainty principle to the shrinkage of
the superconducting wavefunction when $\epsilon$ or $h$ increase, an energy cutoff
must be applied to Eq.~(\ref{summagneto}).\cite{EPLcutoff} This restricts the sum over
$n$ and the integration over $k_{z}$ through $n_{max}=\frac{c-\epsilon}{2h}-1$ and
$|k_{z}^{max}|=\sqrt{c-\epsilon}/\xi_{c}(0)$  (here $c$ is a cutoff constant expected to be of the order of $0.5$),\cite{EPLcutoff,physicaC} leading to
\begin{eqnarray}
&&\Delta\sigma_{ab}=\frac{e^2}{32\hbar \pi\xi_c(0)}\sqrt{\frac{2}{h}}\int_0^{\sqrt{\frac{c-\varepsilon}{2h}}}\mathrm{d}x
\left[\psi^1\left(\frac{\varepsilon+h}{2h}+x^2\right) \right.\nonumber\\
&&\left.-\psi^1\left(\frac{c+h}{2h}+x^2\right)
\right]\,.
\label{magnetoapendix}
\end{eqnarray}
In the zero magnetic field limit, {\it i.e.}, for $h\ll \epsilon,c$, this equation is transformed into
\beq
\Delta\sigma_{ab}=\frac{e^2}{16\hbar \pi \xi_c(0)}\left(\frac{\arctan{\sqrt{\frac{c-\varepsilon}{\varepsilon}}}}{\sqrt{\varepsilon}}-\frac{\arctan{\sqrt{\frac{c-\varepsilon}{c}}}}{\sqrt{c}}     \right)
\label{paracutoff}
\eeq
that corresponds to the paraconductivity under an energy cutoff. At low reduced temperatures and magnetic fields, for $h,\epsilon \ll c$, the cutoff effects become unimportant.\cite{EPLcutoff,physicaC} Accordingly, in this regime Eqs.~(\ref{magnetoapendix}) and (\ref{paracutoff}) reduce to the $c$-independent expressions
\begin{eqnarray}
\Delta\sigma_{ab}=\frac{e^2}{32\hbar \pi\xi_c(0)}\sqrt{\frac{2}{h}}\int_0^{\infty}\mathrm{d}x\,
\psi^1\left(\frac{\varepsilon+h}{2h}+x^2\right)
\label{6e}
\end{eqnarray}
and, respectively, the AL paraconductivity in a $3D$ anisotropic superconductor [Eq.~(\ref{AL})].\cite{AL}
Note finally that Eqs.~(\ref{magnetoapendix}) or (\ref{paracutoff}) lead to the $\Delta\sigma_{ab}$ vanishing at $\varepsilon=c$.

\end{document}